\documentclass[12pt,a4paper]{article}
\usepackage{amsmath,amssymb}
\usepackage{graphicx}
\usepackage{cite}

\allowdisplaybreaks[3]

\begin{document}
\begin{flushright}
KEK-TH-2505
\end{flushright}


\begin{center}
{\Large\textbf{Random Walk in the Boundary and\\
Slow Roll in the Bulk}}
\end{center}

\begin{center}
Yoshihisa \textsc{Kitazawa}$^{2),3)}$
\footnote{E-mail address: kitazawa@post.kek.jp} 
\end{center}

\begin{center}
$^{2)}$
\textit{KEK Theory Center, Tsukuba, Ibaraki 305-0801, Japan}\\
$^{3)}$
\textit{Department of Particle and Nuclear Physics}\\
\textit{The Graduate University for Advanced Studies (Sokendai)}\\
\textit{Tsukuba, Ibaraki 305-0801, Japan}
\end{center}

\begin{abstract}
 The slow rolling inflation is dual to the random walk of conformal zero-mode.
The $O(N)$ enhancement of the two point function of conformal mode $<\omega^2>: N$ is the e-folding number,
suppresses the slow roll parameters by $O(N)$.
The distribution function of conformal mode $\rho_t(\omega)$
 satisfies the Fokker Planck equation.
Under the Gaussian approximation, FP equation boils down to a solvable first order partial differential equation (GFP). 
The identical equation is obained by the thermodynamic arguments.
We study two types of the  solutions of GFP:(1) UV complete spacetime and (2) inflationary spacetime with power potentials. 
The concavity of entangled entropy ensures the potential for inflaton is also concave.
The maximum entropy principle favors the scenario: The universe is (a) born with small $\epsilon$
and (b) grows large by inflation in the concave potentials. 
We predict $1-n_s\geq 0.02(0.016) $ and $r \leq 0.08(0.066)$ for $N=50(60)$ at the pivot angle $0.002Mpc^{-1}$. 
We propose a scenario to produce the curvature perturbation in the right ball park.
\end{abstract}



\newpage

\section{\label{sec:level1}Introduction}
\setcounter{equation}{0}

The important questions on quantum gravity are connected with their Hubble scales.
We may raise the cosmological constant problem, inflation \cite{Infst,InfAG,InfLD,InfPS}, 
dark energy\cite{RP,CDS} and so on .

It is persuasive to relate them to the infra-red behavior of quantum gravity just like hadron physics
with the infra-red behaviors of QCD. With the accumulating observational information, the time is ripe to
make a substantial progress on quantum gravity.   Consistent quantum gravity such as string theory and matrix models are manifestly UV finite. What is difficult is to understand their infra-red behavior. 

In the inflation theory, the universe undergoes the accelerated expansion. 
We build on the stochastic picture of infrared (IR) fluctuations\cite{Starobinsky1986},
\cite{Tsamis2005}.
It is  the diffusion process of  the  conformal zero-mode
like that of the Brownian particles.
The resummation of IR logarithmic effects in quantum gravity, 
which are $O(\log ^na)$ in terms of the scale factor $a$ of the universe 
play an essential role in elucidating cosmic random walk.\cite{Tsamis1994,Kitamoto2012}. 

In our study of IR effects in quantum gravity, Einstein gravity plays the central role as the mass-less sector of string theory contains Einstein gravity. Some modifications of Einstein gravity are proposed in connection with
inflation and dark energy. Phenomenologically it is necessary to introduce a single scalar field (inflaton)
which couples minimally to gravity with a potential. By satisfying a slow roll condition, inflaton can explain
the essence of inflation . Although it is a semi-classical modification of the Einstein gravity,
the quantum effect exists at the heart of the theory, i.e. the vacuum fluctuation is stretched beyond Hubble scale by  inflation.
It should be possible to construct fully quantum theory by  superseding these semi-classical theories.
We argue de Sitter duality is the key to unify  stochastic (quantum fluctuation) and dissipative (entropy generation) aspects.


In order to make this paper self-contained, we write down gravitational  propagators in de Sitter space. 
The de Sitter background is given by 
\begin{align}
ds^2=a_c^2(-d\tau^2+dx_i^2),\hspace{1em}a_c=\frac{1}{-H\tau}.  
\end{align} 
We dress the classical solution by quantum fluctuations.

\begin{align}
ds^2=e^{2\omega} \hat{g}(e^h)^{\mu}_{~\nu} dx_{\mu}dx^{\nu}.  
\end{align} 
The quadratic terms in the Einstein-Hilbert action are given by  
\begin{align}
&\frac{1}{\kappa^2}\int d^4x \sqrt{-g}[R-6H^2]\big|_2 \notag\\
=&\frac{1}{\kappa^2}\int d^4x \big[
-\frac{1}{4}a_c^2\partial_\mu h^{\rho\sigma}\partial^\mu h_{\rho\sigma} 
+\frac{1}{2}a_c^2\partial_\rho h^\rho_{\ \mu}\partial_\sigma h^{\sigma\mu} 
+2Ha_c^3h^{0\mu}\partial_\nu h^\nu_{\ \mu}
+3H^2a_c^4h^{0\mu}h^0_{\ \mu} \notag\\
&\space{4.5em}-2a_c^2\partial_\mu h^{\mu\nu}\partial_\nu\omega
-8Ha_c^3h^{0\mu}\partial_\mu\omega 
+6a_c^2\partial_\mu\omega\partial^\mu\omega
-24H^2a_c^4\omega^2 \big]. 
\label{EH}\end{align}
We adopt the following gauge fixing term: 
\begin{align}
\int d^4x\mathcal{L}_\text{GF}
&=\frac{1}{\kappa^2}\int d^4x\big[-\frac{1}{2}a_c^2F_\mu F^\mu\big], \notag\\
F_\mu&=\partial_\rho h^\rho_{\ \mu}-2\partial_\mu \omega
+2Ha_c h^0_{\ \mu}+4Ha_c\delta_\mu^{\ 0}\omega. 
\label{GF}\end{align}

The sum of (\ref{EH}) and (\ref{GF}) is given by 
\begin{align}
&\frac{1}{\kappa^2}\int d^4x\sqrt{-g}[R-6H^2]\big|_2+\int d^4x\mathcal{L}_\text{GF} \notag\\
=&\frac{1}{\kappa^2}\int d^4x\big[
a_c^2(-\frac{1}{4}\partial_\mu \tilde{h}^{ij}\partial^\mu \tilde{h}^{ij} 
+\frac{1}{2}\partial_\mu h^{0i}\partial^\mu h^{0i}
-\frac{1}{3}\partial_\mu h^{00}\partial^\mu h^{00} 
+4\partial_\mu\omega\partial^\mu\omega) \notag\\
&\hspace{4.5em}+H^2a_c^4(h^{0i}h^{0i}-h^{00}h^{00}+4h^{00}\omega-4\omega^2) \big], 
\end{align}
where $\tilde{h}^{ij}$ is the spatial traceless mode: 
\begin{align}
\tilde{h}^{ij}\equiv h^{ij}-\frac{1}{3}h^{00}\delta^{ij}. 
\end{align}
The quadratic action is diagonalized as follows 
\begin{align}
\frac{1}{\kappa^2}\int d^4x\big[
&-\frac{1}{4}a_c^2\partial_\mu \tilde{h}^{ij}\partial^\mu \tilde{h}^{ij} 
+\frac{1}{2}a_c^2\partial_\mu X\partial^\mu X \notag\\
&+\frac{1}{2}a_c^2\partial_\mu h^{0i}\partial^\mu h^{0i}+H^2a_c^4h^{0i}h^{0i} 
-\frac{1}{2}a_c^2\partial_\mu Y\partial^\mu Y-H^2a_c^4 Y^2 \big], 
\label{diagonalized}\end{align}
where $X$ and $Y$ are given by 
\begin{align}
X\equiv 2\sqrt{3}\omega-\frac{1}{\sqrt{3}}h^{00},\hspace{1em}Y\equiv h^{00}-2\omega. 
\end{align}

In our parametrization\cite{KKK}, the massless scalar field with the canonical kinetic term is
\begin{align}
X=2\sqrt{3}\omega-{1\over \sqrt{3}}h^{00}.
\end{align}
We may retain  massless minimally coupled $X$ field space:
$h^{00}\sim 2\omega \sim (\sqrt{3}/2)X$.
The curvature perturbation $\zeta$
is the fluctuation of the over all scale of the spacial metric:
 $g_{ij}=e^{2\zeta}\hat{g}_{ij}$.  Indeed we find they are the same 
\begin{align}
2\zeta= 2\sqrt{3}\omega+{1\over \sqrt{3}}h^{00}=2X.
\end{align}

We find no IR logarithms in the BRS trivial sector.
As seen in (\ref{diagonalized}) , 
the Einstein gravity consists of massless minimally coupled modes, 
and conformally coupled modes.   In the latter case, the coefficient
of the second de Witt-Schwinger expansion of the propagator cancels
as $R/6-2H^2=0$.
We neglect the conformally coupled modes 
and focus on the subspace of massless minimally coupled modes 
\begin{align}
h^{00}\simeq 2\omega \simeq \frac{\sqrt{3}}{2}X,\hspace{1em}
\tilde{h}^{ij}.
\end{align}

We have computed the one loop quantum correction to Einstein gravity 
\cite{Kitamoto2019-2}.
The bare Einstein gravity is 
inflation theory since it requires the
inflaton field to construct the counter terms.
Einstein gravity is dual to inflation theory in this sense. 
The opposite is true as we obtain the former by integrating out the inflaton from the latter.

In section 2, our key observation in this paper is explained. The duality between the quantum effects of 
Einstein gravity and slowly rolling inflaton holds as the enhancement of the two point function 
leads to the suppression of the slow roll parameters.

In section 3, we argue the random walk of conformal zero mode is dual to slow roll inflation:
This intuition has been our main impetus to pursue dS duality .
In section 4, 
we derive the GFP from classical thermodynamics for inflation theory in a dual picture  to the original quantum stochastic picture. It is based on the relation
$S_A=S_B$. The bulk and boundary entropy are the same.
It justifies  de Sitter duality at the non-perturbative level.
In section 5, we conclude this paper  by listing our successes so far and tasks remaining ahead.
Duality between random walk and slow roll defines the universal class in inflation theories.
The slow roll parameters are $O(1/N)$ and entropy favors the linear potential. If the de Sitter entropy is quantum :EE,
the inflaton potential is concave. 
We construct the composite spacetime in which the curvature perturbation comes out naturally of right magnitude.

\section{de Sitter Duality: Slow Roll versus Stochastic}

\setcounter{equation}{0}

The gravity has always been investigated from geometric point of view following Einstein. Inflation theory is no exception.
On the other hand, the inflation theory leaves too many unexplained freedom. What is the inflaton? Why they perform slow roll on the potential? What determines the potential? and so forth.
There is an alternative (dual) approach, i.e. stochastic point of view. This approach relies on a different  point of view from classical geometric picture.
 The  stochastic equations  on the boundary play the major role like the Einstein's field equation.
This dual approach is capable to explain the essence of inflaton. Namely, the mysterious slow roll can be interpreted  as  the random walk of conformal zero mode in a dual picture.
This idea enables us to predict that  
the magnitude of slow roll parameters as $\epsilon \sim 1/N$.
This universal prediction is explained by the Langevin equation as we recall the 
nutshell of the argument as follows\cite{Parisi}.

At large times, we assume the distribution should be canonical.
\begin{align}
\dot{x}=b(t).
\end{align}
In Einstein gravity, there is only one massless scalar field as explained in the introduction.
So this assumption holds.
Let us consider 
\begin{align}
B(t)^{\delta}=\int_t^{t+\delta}dt' b(t').	\end{align}
We assume $\delta$ can be taken arbitrary small.  
In the low energy approximation, we can ignore the drift term.

Under such circumstances, 
\begin{align}
\overline{b(t_1)b(t_2)}=\delta(t_1-t_2)H\tilde{g}.
\label{Lancor}
\end{align}
$\tilde{g}=(3/4)g$ is the coupling projected in $X$ field.
The variance is computed as
\begin{align}
\overline{(x(t)-x(0))^2}=\int_0^tdt_1\int_0^tdt_2\overline{b(t_1)b)(t_2)}
=\int dN\tilde{g}.
\label{Lang}
\end{align}
The scalar spectral index is   $\tilde{g}(t)/\int^t dN\tilde{g} \sim 1/N$ if
the slow roll conditions are satisfied $1>>1/N$.
It is clear that  
the tensor to scalar ratio  is  $ {g}(t)/\int^t dN\tilde{g} \sim 1/N$  anagolously.
The  slow roll parameters  for the power potential models will be
worked out in the next section.

The probability distribution of $x(t)-x(0)$ is
\begin{equation}
P(x,0)= \sqrt{ \frac{1}{2\pi\tilde{g}N}}\exp[-\frac{(x-x(0))^2}{2\tilde{g}N}].
\end{equation}
This is the diffusion kernel for constant $\tilde{g}$. It remains a good approximation for slow roll models.
\begin{align} 
 \frac{\partial}{\partial N}P(x,0)= \frac{1}{2}\tilde{g}\frac{\partial^2}{\partial^2 x}P(x,0)
 \label{GFP1}
\end{align}
The Langevin equation (\ref{Lang}) is equivalent to the Fokker Planck equation (\ref{GFP1})
as shown below ,
\begin{equation}
\frac{\partial}{\partial N}<\omega^2>= \int d\omega \omega ^2  \frac{\partial}{\partial N}P
=\int d\omega \frac{1}{2}\tilde{g} \omega ^2\frac{\partial^2}{\partial^2 \omega ^2 }P(\omega,0)
=\tilde{g}.
\end{equation}
They are the equations in $1+1$ dimensions:$(N,\omega )$. de Sitter duality implies the 4 dimensional slow roll inflation
theory is equivalent to 2 dimensional FP theory. Since $4=2+2$, this is the holographic correspondence
expected in quantum gravity.

We have thus far pointed out the duality between the slow roll gravitational models and FP equations with no gravity.  The logic behind it is the de Sitter duality between 
Random walk and slow roll inflation.  The two point functions  of respective theories display
the identical scaling law for large $N$ which indicates they belong to the same
universal class.
 The duality holds after resumming IR logs to all orders.
In fact the 2 point function gives the  logarithmic contributions
$(g \log (a))^n\sim (gN)^n$. In order to determine the cosmological implications,
we need to resum them to all orders.

After the resummation,
we obtain FP equation from the renormalization group.
It can determine the evolution of the entropy of the universe.
So each solution of FP equation corresponds to a history of the universe.  

In de Sitter space, the scale invariance of the infra-red quantum 
fluctuations may give rise to time dependent effects such as shielding the cosmological constant. 
While the inflaton generates entropy by dissipation, it represents the quantum effect in de Sitter dual picture in accordance with the fluctuation-dissipation  theorem. The leading log contributions
$(gHt)^n\sim (g\log a)^n$ grow during the long cosmic history.
The renormalization group is one of the most powerful tool to accomplish it.
In the critical phenomena, the divergence of the correlation length is 
caused by IR degrees of freedom.
The important point is that there is only one massless scalar mode  which
plays the significant role as explained in the introduction. 
We call it  $X$ field. It is nothing but the curvature perturbation $\zeta$.

The remarkable relation $S_A=S_B$ holds when we consider $A=S^3, B= H^4$
as the boundary-bulk pair.
The entropy of bulk (sub-horizon) and boundary (super-horizon) is the same.
This fact indicates the existence of de Sitter duality. 
It may explain that the FP equation possesses  UV finite solutions.
It further indicates that 
the Universe starts with the de Sitter expansion near the Planck scale with 
$\beta\sim\epsilon=0$. 

In order to elucidate the IR logarithmic effects non-perturbatively, 
we formulate a Fokker Planck equation for the conformal zero  modes of 
the metric. 
We obtain the $\beta$ function for dimension-less gravitational
coupling $g=G_NH^2/\pi$ in a Gaussian approximation . $G_N$ is the Newton's coupling and $H$ is the Hubble parameter. It is an excellent approximation in quantum gravity  for a small coupling $g$ except at the beginning of the Universe. 
We first sum up the $O(\log ^na)$ terms to all orders to identify the
one-loop running coupling $g=2/\log N$.
We the $1/(\log ^nN)$ type corrections to  take account of  the quantum back-reaction on $g$. 
Since the $\beta$ function turns out to be negative, 
it implies asymptotic freedom for $g$ toward the future
\cite{Gross1973,Politzer1973}. 
Furthermore, the $\beta$ function possesses the ultraviolet (UV) fixed point in the past with the critical coupling $g=1/2$. 
This fact indicates that our Universe begun with the dS expansion near the Planck scale with a minimal entropy.

Since $\epsilon\propto \beta$, the  smallness of
$\epsilon$ in the beginning of the Universe is naturally explained. 
In our approach,
the quantum correction to de Sitter entropy is given by the Gibbs entropy of the zero mode. 
We may decompose the total Hilbert space  $H=H_A\otimes H_B$ into   $H_A$ and $H_B$. 
Quantum definition of $\rho_B$ involves the integration on A: $ \rho_B=Tr_A\rho^{tot}$.
This operation corresponds to the renormalization of $\rho$ at the horizon scale.
Then 
$S_B=-Tr_B\rho_B\log(\rho_B)$.
We have counted the entropy of the mass less sector
 in the super-horizon. 
  At the quantum level, the de Sitter entropy is von Neumann entropy, i.e.  entangled entropy: EE.
  
  \section{Conformal mode in the leading log}

\setcounter{equation}{0}

The solution of FP equation implies the conformal mode performs the
random walk.
Since the fractal dimension of random walk $D_f=2$ is universal,  
the 2 point function is $O(N) $.
We can sum up the IR logarithms $\log ^no$ by this equation to find 
a running coupling $g(t)$. 
We first work out the leading log solution which 
 is valid in the large $N$ limit.
The FP (diffusion) equation  shows that the solution is the Gaussian
distribution with the  standard deviation  $O(1/N(t))$ as shown in section 2 \cite{Parisi}. 
 The distribution entropy of the conformal mode  increases logarithmically, 
\begin{align}
\frac{1}{2}\log \frac{1}{\xi}
\sim\frac{1}{2}\log N(t)
\label{entropy}\end{align} 
In de
Sitter space, entropy may be identified with the effective action 
with the opposite sign. 
$-\Gamma=S$ as $E=0$.
We maximize the entropy in order to minimize the effective action.

Identifying the EE of conformal zero mode
with the quantum correction to dS entropy,
we obtain the bare action with the counter term
\begin{align}
\frac{1}{g_B}=\frac{1}{g(N)} - \frac{1}{2}\log(N). 
\label{coupling}\end{align}
By requiring the bare action is  
independent of the renormalization scale: namely  $N$ , as we have just 
done to derive FP equation,
we obtain the one loop $\beta$ function.
\begin{align}
\beta=\frac{\partial}{\partial \log(N)}g(N)=-\frac{1}{2}g(N)^2. 
\label{beta1}\end{align}
We find the running gravitational coupling in the leading log approximation:
\begin{align}
 g(N)={2/ \log(N)}, 
\end{align}
while
 the von-Neumann entropy of the conformal mode  increases logarithmically, 
\begin{align}
S(N)={1/ g(N)}={\log(N)/ 2}. 
\label{beta11}\end{align}
 We regard these facts as the semiclassical evidence that de Sitter entropy is EE. 
 
 We have shown the renormalized distribution function obeys gravitational Fokker-Planck equation (GFP). 
That is equivalent to tune $\xi$ self-consistently. Under the Gaussian approximation, they boil down to a solvable first order partial differential equation. 
The identical equation are derived by the thermodynamic arguments in the inflationary space-time. GFP 
 determines the evolution of de Sitter entropy of the universe. 
 It coincides with  $1/g$.

The holographic investigation at the boundary shows that 
$g$ is asymptotically free toward the future. 
The renormalization group trajectory must reach Einstein gravity in the weak coupling limit 
for the consistency with general covariance \cite{Kawai1993}. 
We find that it approaches a flat space-time in agreement with this requirement.

We have evaluated the time evolution of EE entropy to the 
leading log order in \cite{Kitamoto2019-2}.
We reproduce these arguments as precise as possible here.  
In order to take account of the higher loop corrections in $g$, the FP equation 
should be uniquely generalized. 
It turns out to be just necessary to make the equation covariant and local:
\begin{align}
\frac{\partial}{\partial N}\rho(N)
-\frac{3g(N)}{4}\cdot\frac{1}{2}\frac{\partial^2}{\partial \omega^2}\rho(N)
=0.
\label{GFP}\end{align}
The covariance and locality ensure that
 the all the  $n$-th loop divergences are local 
if  all the sub divergence are subtracted.
Furthermore the bare operator does not depend on the renormalization point (time).

In what follows, we adopt the following Gaussian anzats of FP equation.
\begin{align}
\rho=\sqrt{\frac{4\xi (t)}{\pi g(t)}}\exp\big(-\frac{4\xi (t)}{g(t)}\omega^2\big).
\label{gausaz}
\end{align}

The standard deviation $\xi \propto 1/N$ decreases as we pointed out in the preceding section.

This distribution function determines the two point function of the conformal zero mode.
\begin{align}
\int d\omega \omega^2 \rho(\omega) ={g(t)/ 8\xi(t)}.
\end{align}
 We put the Gaussian ansatz into the FP equation (\ref{GFP1}) and
find the following condition for the background to satisfy.
\begin{align}
{\partial / \partial N} 
\left({g/ 8\xi}\right)={3g/ 4}.
\end{align}

We thus obtain a  simple equation in terms of e-foldings.
\begin{align}
\frac{\partial}{\partial N}\log \frac{g(t)}{\xi}=6\xi. 
\label{FP3}\end{align}
This is the equation to investigate in what follows . Since the  EE is $(1/2)\log \frac{g(t)}{\xi} $,
this equation determines the evolution of entropy of a universe.
This equation is derived from FP equation with the Gaussian approximation.
The same equation holds in the inflation theory as explained later.
In this way (\ref{FP3}) supports de Sitter duality between Stochastic
and thermodynamic approaches.

 We have investigated FP equations from various view points. We have first adopted the microscopic approach. 
 We derive FP equation from renormalization group. We then derive it from macroscopic thermodynamics. 
 We have also checked the consistency with Langevin formalism.

We first investigate a two  point function of the conformal mode in the
coincident time limit $<{\omega}^2>$ as it is a subtle step.

We recall the following identity holds at the horizon exit $t=t^*$\cite{Maldacena203}.
\begin{align}
\dot{N}^*\exp({N^*})=k.
\label{IRlog}
\end{align}
where $*$ denotes the exit time.
 It is the reason why the renormalization scales are 
 related $\log k = Ht^*$ in
 (\ref{IRlog}). 
 It is the first come, the first served system. 
 The softer the plain wave, the earlier its exit time.

We need to use time independent UV cut-off like dimensional regularization to identify time dependent quantum
corrections. 
This is nothing but the second term of the Schwinger-de Witt expansion of the propagator.
It is logarithmically divergent near 4 dimensions as $1/\delta\sim 1/4-d$ pole indicates 
\begin{align}
<\omega^2>=-\frac{3}{16\pi}
 \frac{G_N{R}}{ 6}(\frac{2}{\delta}
 -{2}{\log(k_*)})
 \label{SDT}
 \end{align}
Although the leading term is  quadratically divergent,
 it vanishes in the dimensional regularization. 
As we emphasize, there is no time dependent UV contributions. 
We focus on the Hubble scale physics where $a(t)=1/(-\tau H)=\exp(Ht)$
and $|k\tau|\sim 1$. Our renormalization scale is
$\log k\sim-\log (-\tau )\sim Ht=N$.
Note the metric is negative,
but  $k$ is the exiting momenta at the horizon and it increases with exiting time $t^*$.
The renormalized 2 point function is

\begin{align} 
 <\omega^2>_R=
 ({3/ 4})g^*\log(k_*)
=({3/ 4})g^* N^*
\end{align}

The bare distribution function is
\begin{align}
\rho_B= \exp (\tilde{g}^*({1/ \delta}-\log(k_*)){\partial^2 / \partial \omega ^2}) \rho(\omega)
\end{align}

The FP equation is obtained as $\rho_B$ is independent of the renormalization time.
\begin{align}
 \dot{\rho}= \tilde{g}{\partial^2 / \partial \omega^2} \rho(\omega)
\end{align}

While the wave functions of the bulk modes oscillate with respect to $\tau$,
the boundary mode is constant. That is why we call it the zero mode.
Our strategy is to integrate out oscillating modes first.

We thus construct low energy effective 
theory around the Hubble scale by renormalizing the distribution function.
This is the one of the fundamental results of our work
\cite{Whaleboat}. We investigate it in the both
its foundation and implications in this work.
Our theory is identical to that of Brownian particles in $1+1$ dimensions.
It is holographic in comparison to $dS_4$. In our theory 
conformal zero-mode random walk on the boundary (horizon).

As we have emphasized, Gaussian approximation must be valid unless 
we probe Planck scale.
(\ref{FP3}) determines the evolution of EE  $S={1/2}\log{g/ \xi}$ with respect to $N$.
This formula  confirms the validity of our postulate that distribution entropy
of conformal zero mode constitutes the quantum correction to de Sitter entropy
\cite{correction}.

Secondly, the leading log approximation is valid in the large $N$ limit
since $g\sim O(\log N)$ can be regarded as a constant.
We obtain the equation which is the large $N$
limit of (\ref{FP3}).
Equivalently we put it as follows.
\begin{align}{\partial/ \partial N}{\log \xi}=-6\xi.
\label{FP4}\end{align}

The solution is
\begin{align}
\xi=\frac{1}{6N}. 
\label{oxisol}\end{align}
Here we have neglected time dependence of $1/g \sim \log N$
in comparison to $1/\xi\sim N$.
In such a limit, (\ref{FP3}) turns into
(\ref{FP4}). 
For finite $N$, $\xi=1/6N$ will acquire finite $N$ corrections like
 (\ref{solution1}).

Since $\epsilon=-(1/2)({\partial / \partial N})\log g$,
 it happens $\epsilon=1/4N$ for the linear potential, the result is in agreement with the slow roll picture $<\zeta^2> =g/\epsilon$. In this estimate, we have used Gaussian ansatz and large $N$
expansion. It appears that only concave potential is allowed: $n>1$ if the de Sitter entropy is EE.

The solutions with more generic potentials are considered in the next section from  de Sitter duality point of view.
Our basic conjecture is that the mysterious slow roll of inflaton represents the Brownian motion of the conformal zero mode. 
This idea explains generic features and may develop  further to demystify  de Sitter duality.
The attractive point of our theory is its simplicity. 
The inflaton is replaced by the random walk of the conformal zeromode. The universal enhancement of scalar to tensor ratio by $O(N)$ is explained by its fractal dimension of the scalar modes. 
However the precise coefficient depends on the details.

There is a UV fixed point in our renormalization group.
FP equation (\ref{FP3}) enables us to evaluate higher order corrections to the $\beta$ function. 
The expansion parameter is $1/\log N$. 
We can confirm that the following $g_f$ and $\xi_f$ satisfies (\ref{FP3}), 
\begin{align}
g_f=\frac{2}{\log N}\big(1-\frac{1}{\log N}\big),~~
{\xi}_f=\frac{1}{6N}\big(1-\frac{1}{\log N}\big). 
\label{solution1}\end{align}
The leading log results are (\ref{beta1}) and (\ref{oxisol})  respectively.

Thus, the $\beta$ function, $\epsilon$ and the semi-classical entropy generation rate are given by 
\begin{align}\beta={\partial g}/{\partial \log N}=
 -{2/ \log^2 N}+{4/ \log(N)^3}.
\label{beta2}\end{align}
\begin{align}
\epsilon_f = -({1/ 2}){\partial \log(g_f)/ \partial N}=
 -{1/ 2g_f N}\beta_f.
 \end{align} 
\begin{align}
\frac{\partial}{\partial N}S_{cc}=\frac{\partial}{\partial N}\frac{1}{g_f}
=-\frac{1}{Ng_f^2}\beta_f.
\label{entropy2}\end{align}

A remarkable feature is that the coupling has the maximum value $g=1/2$ at the beginning. 
It steadily decreases toward the future 
as the $\beta$ function is negative in the whole region of time flow. 
It has two fixed points at the beginning and at the future of the Universe. 
The existence of the UV fixed point may indicate the consistency of quantum gravity. 
The single stone solves the
$\epsilon$ problem \cite{Penrose1988} as well since it vanishes at the fixed point. 
The $\beta$ function describes a scenario that 
our Universe started the dS expansion with a minimal entropy $S=2$ while it has $S=10^{120}$ now. 

Since we work with the Gaussian approximation, our results on the UV  fixed points are not water tight
as the coupling is not weak. Nevertheless we find it remarkable that they
support the idea that quantum gravity has a UV fixed point 
with a finite coupling. In fact 4 dimensional de Sitter space is constructed in the target space 
at the UV fixed point of $2+\epsilon$ dimensional quantum gravity  \cite{Kawai1993}.
4 dimensional de Sitter space also appears at the UV fixed point of the exact
renormalization group \cite{Reuter}\cite{Souma}.

Such a theory might be a strongly interacting conformal field theory .
However, it is not an ordinary field theory as the Hubble scale is Planck scale. 
Our dynamical $\beta$ function is closely related to the cosmological horizon and physics around it. 
The existence of the UV fixed point could solve the trans-Planckian physics problem. 
A consistent quantum gravity theory can be constructed 
under the assumption that there are no degrees of freedom at trans-Planckian physics \cite{Bedroya2019}. 
In this sense, it is consistent with string theory and matrix models.  
The Universe might be governed by (\ref{solution1}) in the beginning 
as it might be indispensable to construct the UV finite solutions of the FP equation.

 \section{Inflationary Universe with Power Potentials}
\setcounter{equation}{0}

The inflaton may be identified with the stochastic variable $f$ whose correlators show characteristic features of Brownian motion.  $<f^2>=\tilde{N}$ and $g\propto \tilde{N}^{m/2}=<f^m>$. 
The increase of the entropy $S=1/g$ can be evaluated by the first law $T\Delta S=\Delta E$ 
where $\Delta E$ is the incoming energy flux of the inflaton  \cite{Frolov2003}.
Obviously we are estimating the entropy of super-horizon.

Since these solutions are scale invariant, Langevin
equation determines the scaling behavior precisely
in contrast to the logarithmic correlators. correlators

In this way, the one of the Einstein's equation is obtained:
\begin{align}
\dot{H}(t)=-4\pi G_N \dot{f}^2.
\end{align}
This formula can be re-expressed 
in terms of the slow roll parameter: 
\begin{align}
\epsilon=\dot{f}^2/2H^2
=-(1/2)\frac{\partial }{ \partial {N}}\log g(t).
\label{epxi}
\end{align}

The tilt of the gravitational wave spectrum is :
$n_t=-2\epsilon$.
It is the one loop renormalization effect $g\propto \exp(-2\epsilon N) \sim a_c^{-2\epsilon}$
 \cite{Kitamoto2019-2}  \cite{BCFH}.
However we need to resum the IR log terms to all orders as explained in section 3.

We add $O(1/\tilde{N})$ quantity to the both sides of the equation.
\begin{align}
2\epsilon+\frac{1}{ \tilde{N}}=-\frac{\partial} { \partial {N}}\log (g \tilde{N}).
\label{tiltomega}
\end{align}

We find the relation between  $\xi$ and $\epsilon$
in  : 
\begin{align}6\xi=2\epsilon+\frac{1}{\tilde{N}}=\frac{(m+2)}{2\tilde{N}}\end{align} 
is reproduced.
The tilt of the conformal mode is reproduced.
We have thus shown that  curvature and conformal perturbation are identical up to the constant factor
(\ref{solution2}).
For power potential inflationary universe, (\ref{mf2}) is rewritten as

\begin{align}
6\xi=\frac{\partial}{ \partial {N}}\log \frac{g}{ \xi}
=\frac{\partial }{ \partial {N}}\log \frac{g}{ \epsilon}\end{align}.

It is consistent with our GFP (\ref{FP3}). The second equality arises since $\xi/\epsilon$ is independent of ${N}$ as shown in (\ref{epxi}). This ambiguity
corresponds to a constant $c$ in (\ref{solution2}). 
In other words,  
$\xi=\frac{m+2}{12\tilde{N}}, \epsilon=\frac{m}{4\tilde{N}}$ are $O(1/\tilde{N})$. The both can balance the equation.

In this form, it is evident that $6\xi$ is the scalar spectral index in agreement with slow roll inflation theory. Tensor to scalar ratio is $16\epsilon$ .
We have shown GFP can be derived by macroscopic arguments also.
From quantum gravity point of view, this is an important non-perturbative evidence for 
de Sitter duality: quantum/geometric duality.

EE must be concave  from strong sub-additivity\cite{HolographicEE}\cite{EEinQFT}.
Inflationary universe with a slow roll parameter may be well approximated  as de Sitter space locally

The concave power solutions can be obtained from convex solutions by formally replacing $m$ by $1/n$ where
$n>1$ has to be a real number .

 \begin{align}
 \frac{\partial^2}{ \partial N^2}(\frac{1}{ g})=\frac{(1/n)(1/n-1)}{N^2}(\frac{1}{ g}),
~~
\frac{\partial^2 }{\partial N^2} S<0
\end{align}

We conclude that potentials for consistent inflation theory are concave.
In fact convex potentials are disfavored by observation.

The conformal mode of the inflationary universe is performing the Brownian motion.
Its trajectories constitute  2 dimensional  fractal.
The concave potentials are promising avenue to explore right now.
The potential of inflation theory may be identified with the entangled entropy
which is always concave.

As it turns out, the equation (\ref{FP3}) has another class of solutions:
the inflationary universe  with the power potentials.
\begin{align}
g=c\tilde{N}^\frac{m}{2},\hspace{1em}\xi=\frac{m+2}{12\tilde{N}}.
\label{solution2}\end{align}
Since these are solutions of FP equation, they have many common features. 
The enhancement of scalar 2 point function by $O(N)$ is such an example.

Here $c$ is an integration constant.
$m$ denotes the power of the potential: $f^m$.
We have changed the variables by
replacing $N$ by $\tilde{N}$
where $\tilde{N}=N_e-N$.  $N_e$ denotes the e-folding number at the end of inflation.  To be precise, they are the solutions of the following equation:
\begin{align}
-\frac{\partial}{\partial N}\log \frac{g(\tilde{N})}{\xi(\tilde{N})}=6\xi(\tilde{N}). 
\label{FPtr}\end{align}

 (\ref{solution2}) shows  $\epsilon$ and $\xi$ are related as 
\begin{align}
\epsilon=\frac{m}{4\tilde{N}}=\frac{3m}{m+2}\xi.
\label{epxi}
\end{align}

In this way, the one of the Einstein's equation is obtained:
\begin{align}
\dot{H}(t)=-4\pi G_N \dot{f}^2.
\end{align}
This formula can be re-expressed 
in terms of the slow roll parameter: $\epsilon={\dot{f}^2}/2H^2$
\begin{align}
2\epsilon=-\frac{\partial }{ \partial {N}}\log g(t).
\end{align}
The tilt of the gravitational wave spectrum is $n_t=-2\epsilon$.
We add $O(1/\tilde{N})$ quantity to the both sides of the equation.
\begin{align}
2\epsilon+\frac {1}{\tilde{N}}=-\frac{\partial } {\partial {N}}\log (g \tilde{N}).
\label{mf2}
\end{align}
We find the relation between  $\xi$ and $\epsilon$
in (\ref{epxi}) : 

\begin{align}6\xi=2\epsilon+\frac{1}{\tilde{N}}=\frac{(m+2)}{2\tilde{N}}\end{align} 
is reproduced.
We have thus shown that  curvature and conformal perturbation are identical up to the constant factor
(\ref{solution2}).

For power potential inflationary universe, (\ref{mf2}) is rewritten as
\begin{align}
6\xi=\frac{\partial }{ \partial {N}}\log {\frac{g}{ \xi}}=\frac{\partial}  {\partial {N}}\log {\frac{g}{ \epsilon}}.
\end{align}
It is consistent with our GFP (\ref{FP3}). The second equality arises since $\xi/\epsilon$ is independent of ${N}$ as shown in (\ref{epxi}). This ambiguity
corresponds to a constant $c$ in (\ref{solution2}). 
In other words,  $\xi={m+2}/{12\tilde{N}}, \epsilon={m/ 4\tilde{N}}$ are $O(1/\tilde{N})$. The both can balance the above equation.

In this form, it is evident that $6\xi$ is the scalar spectral index in agreement with slow roll inflation theory. Tensor to scalar ratio is $16\epsilon$ .
We have shown GFP can be derived by macroscopic arguments also.
From quantum gravity point of view, this is an important non-perturbative evidence for 
de Sitter duality: quantum/geometric duality.


\begin{table}
\begin{center}
\begin{tabular}{ccccc}
&
& $n=1$
& $n=2$
&$n=3$
\\[.5pc] \hline\hline
&$\epsilon $
&$\frac {1}{ 4\tilde{N}}$
& $\frac {1}{8\tilde{N}}$
& $\frac {1}{12\tilde{N}}$
\\[.5pc] \hline
&$n_s$
&$1-\frac{3}{2\tilde{N}}$
&$1-\frac{5}{4\tilde{N}}$
& $1-\frac{7}{6\tilde{N}}$
\\[.5pc] \hline\hline
&$dn_s/dN$
&$-\frac{3}{2\tilde{N}^2}$
&$-\frac{5}{4\tilde{N}^2}$
& $-\frac{7}{6\tilde{N}^2}$
\\[.5pc] \hline
\end{tabular}
\caption{\label{tab:tm1} }

\end{center}
\end{table}

\begin{table}
\begin{center}
\begin{tabular}{ccccc}
& $\tilde{N}=50$
& $n=1$
& $n=2$
&$n=3$
\\[.5pc] \hline\hline
&$r $
&$0.08$
& $0.04$
& $0.027$
\\[.5pc] \hline
&$n_s$
&$0.97$
&$0.975$
& $0.977$
\\[.5pc] \hline\hline
&$dn_s/dN$
&$-0.0006$
&$-0.0005$
& $-0.00046$
\\[.5pc] \hline
\end{tabular}
\caption{\label{tab:tm2} N=50}
\end{center}
\end{table}
In Table \ref{tab:tm1}, we list expected $\epsilon,1-n_s,dn_s/dN$ in the power potential
model with $f^{1/n},n=1,2,3$. $n>1$ for the concave potential.
We note $1-n_s$ is bounded from below by $1/\tilde{N}$ while
$r=16\epsilon $ is not.
Table II lists our numerical estimates by putting $\tilde{N}=60$ in Table I.
\begin{table}
\begin{center}
\begin{tabular}{ccccc}
& $\tilde{N}=60$
& $n=1$
& $n=2$
&$n=3$
\\[.5pc] \hline\hline
&$r $
&$0.066$
& $0.033$
& $0.022$
\\[.5pc] \hline
&$n_s$
&$0.975$
&$0.979$11
& $0.981$
\\[.5pc] \hline\hline
&${dn_s/dN}$
&$-0.0006$
&$-0.0005$
& $-0.00046$
\\[.5pc] \hline
\end{tabular}
\caption{\label{tab:tm3},N=60}
\end{center}
\end{table}

  They are consistent with the current observations \cite{2018}.

   \begin{table}
\begin{center}
\begin{tabular}{cc}
&Planck 2018
\\[.5pc] \hline\hline
&$ r_{0.002}<0.065$
\\[.5pc] \hline
&$n_{s,0.002}=0.979 \pm 0.0041$
\\[.5pc] \hline
&$dn_s/dN = - 0.006 \pm 0.013$
\end{tabular}
\caption{\label{:tm4}, Planck 2018}
\end{center}
\end{table}
We adopt the same pivot scale $0.002Mpc^{-1}$ for $r,n_s$ to respect self-consistency.
It is important to establish the bound on $n$.
We recall here the curvature perturbation:
\begin{align}
P\sim 2.2 \times 10^{-9}.
\label{CP}
\end{align}
It is bounded from below $g>10^{-11}$ if $\epsilon>1/200$.

The entropy grows logarithmically $1/g=(\log N)/2$ as $N$ grows after the Universe emerges out of the UV fixed point.
In the inflationary phase, the entropy grows power like $1/g\sim 1/\tilde{N}^{1/2n}$.
The entropy increases slowly  in the pre-inflation era while
it grows much faster in the inflation era. 
Inflation phase inevitably takes over the pre-inflation phase
as the entropy of the former dominates in time.

The simplest scenario assumes that the inflation starts sometime after the 
birth of the Universe .  We hope to explain  the observed magnitude of the curvature perturbation
$P\sim 10^{-10}$. 
This is a fundamental quantity of the Universe.
 The inflation does not stop unless $\epsilon$ grows to  $O(1)$, it is necessary to  cross over to
the slow roll inflation phase.

We find it attractive to consider the composite solutions, which consist of the logarithmic solution (\ref{solution1}) 
and the the power-law solution (\ref{solution2}).
They may describe the whole history of the Universe from its birth at the UV fixed point to the completion of inflation.
Let us investigate a pre-inflation scenario.
The entropy grows logarithmically $1/g=(\log N)/2$ as $N$ grows after the Universe emerges out of the UV fixed point.
In the inflationary phase, the entropy grows power like $1/g\sim 1/\tilde{N}^{1/2n}$.
The entropy increases slowly  in the pre-inflation era while
it grows much faster in the inflation era.

\section{Conclusions and Discussions}

Although we have 
 argued the necessity of pre-inflation universe, it is another matter to understand the mechanism and the consistency of the transition. 
We have explored  composite solutions which realize the above transition scenario
\cite{Jain2007}.
The system consists of the logarithmic solution (\ref{solution1}) 
and the the power-law solution (\ref{solution2}).  The entropy $1/g$ always increase with cosmic evolution.
They may describe the whole history of the universe from its birth at the UV fixed point until the finale of the inflation.

Let us investigate pre-inflation scenario.
The entropy grows logarithmicaly $1/g=(\log N)/2$ as $N$ grows after the Universe emerges out of the UV fixed point.
In the inflationary phase, the entropy grows power like $1/g\sim 1/\tilde{N}^{1/2n}$.
The entropy increases slowly  in the pre-inflation era while
it grows much faster in the inflation era.
Inflation phase inevitably takes over the pre-inflation phase
as the entropy of the former dominates in time.

The simplest scenario assumes that our universe is the $(50:50)$ composite of the UV complete and inflationary universes. 
We may consider composite universe with the geometrically  averaged
Hubble parameter: $H^2=H_1H_2$. The linear average of $\epsilon$ is
$\bar{\epsilon}=(\epsilon_1+\epsilon_2)/2$.

We recall $\epsilon=-(1/2){\partial \over  \partial N} \log H^2$.
 $\epsilon_1 \sim 1/2N\log(N) , \epsilon_2 \sim 1/4n\tilde{N}$ .
Note the former decreases while the latter increases with $N$.
Therefore the inflationary universe dominates when $|\epsilon_1|<|\epsilon_2|$.
An explicit estimate of the transition point with $n=2$
\footnote{The concavity requires $n>1$. Entropy favors small $n$. } is
$N \sim 2 \times 10^9,  \tilde{N} \sim 8 \times10^9$.
The inflation is terminated at $\epsilon_*=1,\tilde{N}=0.1$.


Our transition scenario is supported by the expectation that the solution with dominant entropy  is chosen. 
Since the entropy $1/g$ for (\ref{solution1}) increases logarithmically 
and that for (\ref{solution2}) increases in a power-law, 
the former is chosen initially and the latter is chosen if we wait long enough.

Holography is a specific feature of quantum gravity.
In this paper we have derived the Fokker-Planck equation
on the boundary from the renormalization group equation 
in the bulk. We have stressed the de Sitter duality between 
the bulk geometry and the boundary  random walk at the horizon.
de Sitter duality is deeply connected with the von Neumann or entangled entropy of the conformal zero-mode.

Suppose the total Hilbert space is factorized into
the bulk states and boundary states. The entropy of the bulk and
the boundary are the same. This explains why de Sitter duality is   important.
It is not expected applioli that $\beta$ function at the UV fixed point
can be accessible from the FP equation at the boundary. 
Nevertheless the dual pair have the identical entropy which
coincide with the  effective action.   
This set-up allows us to evaluate EE in the both quantum procedure and in the classical 
thermodynamics.

Furthermore we can even pull out the promising candidate
for the curvature perturbation. Its  scale is set when the cosmic inflation begun $\tilde{N}\sim 10^{10}$. 
 Its an encouraging sign for the relevance of our  research to quantum gravity on de Sitter space.

\section*{acknowledgment}

This work is supported by Grant-in-Aid for Scientific Research (C) No. 16K05336. 
We thank Hiroyuki Kitamoto,  Masashi Hazumi, Satoshi Iso, Hikaru Kawai, Kozunori Kohri, 
Takahiko Matsubara, Jun Nishimura, Hirotaka Sugawara, Takao Suyama and Yuko Urakawa
 for discussions.

\end{document}